# Phase-merging Enhanced Harmonic Generation Free-electron Laser

**Chao Feng, Haixiao Deng** [1] **, Dong Wang, and Zhentang Zhao**
Shanghai Institute of Applied Physics, Chinese Academy of Sciences, Shanghai 201800, China

E-mail: denghaixiao@sinap.ac.cn

**Abstract.** Together with one of its variants, the recently proposed phase-merging enhanced harmonic generation (PEHG) free-electron lasers (FELs) have been systematically studied in this paper. Different form a standard high-gain harmonic generation scheme, a transverse gradient undulator is employed for introducing a phase-merging effect into the transversely dispersed electron beam in PEHG. The analytical theory of the phase-merging effect and the physical mechanism behind the phenomenon were presented. Using a representative and realistic set of beam parameters, intensive start-to-end simulations for soft x-ray FEL generation were given to illustrate the performance of PEHG. Moreover, some practical issues that may affect the performance of PEHG were also discussed.

## 1. Introduction

The recent success of self-amplified spontaneous emission (SASE) based x-ray free-electron laser (FEL) facilities [1, 2] is enabling forefront science in various areas. While the radiation from a SASE FEL has excellent transverse coherence, it typically has rather limited temporal coherence as the initial radiation comes from the electron beam shot noise. To overcome this problem, several SASE-based techniques have been developed, mainly including self-seeding [3, 4, 5], purified-SASE [6], improved-SASE [7], and HB-SASE [8], etc.

An alternative way for significantly improving the temporal coherence of high-gain FELs is frequency up-conversion schemes, which generally relay on the techniques of optical-scale manipulation of the electron beam phase space with the help of external coherent laser sources. In the high-gain harmonic generation (HGHG) scheme [9], typically a seed laser pulse is first used to interact with electrons in a short undulator, called modulator, to generate a sinusoidal energy modulation in the electron beam at the seed laser wavelength. This energy modulation then develops into an associated density modulation by a dispersive magnetic chicane, called the dispersion section (DS). Taking advantage of the fact that the density modulation shows Fourier components at the high harmonics of the seed, intense radiation at shorter wavelengths can be generated. The output property of HGHG is a direct map of the seed laser's attributes, which ensures high degree of temporal coherence and small pulse energy fluctuations with respect to SASE. These theoretical predictions have been demonstrated in the HGHG experiments [10-13]. However, significant bunching at higher harmonics by strengthening the energy modulation would increase the energy spread of the electron beam, which would result in a degradation of the amplification process in the radiator. The requirement of FEL amplification on the beam energy spread prevents the possibility of reaching short wavelength in a single stage HGHG. In order to improve the frequency multiplication efficiency in a single stage, more complicated phase space manipulation techniques have been developed, e.g., the echo-enabled harmonic generation [14, 15] technique employs two modulators and two dispersion sections, which can be used to introduce echo effect into the electron beam phase space for enhancing the frequency multiplication of the current modulation with a relatively small energy modulation.

---
[1] To whom any correspondence should be addressed.

Recently, a novel phase space manipulation technique, originally named as cooled-HGHG, has been proposed for significantly improving the frequency up-conversion efficiency of harmonic generation FELs [16]. This technique benefits from the transverse-longitudinal phase space coupling, while other harmonic generation schemes only manipulate the longitudinal phase space of the electron beam. When the transversely dispersed electrons pass through the transverse gradient undulator (TGU) modulator, around the zero-crossing of the seed laser, the electrons with the same energy will merge into a same longitudinal phase, which holds great promise for generating fully coherent short-wavelength radiation.

At the first glance, this phase-merging phenomenon is very similar with the electron beam energy spread cooling. However, the beam energy spread within the range less than seed laser wavelength is reduced, while the global beam energy spread does not change in such a process. Therefore, in order to clearly and unanimously illustrates the physics behind it, we rename such a scheme as phase-merging enhanced harmonic generation (PEHG), although further studies demonstrate that, this novel technique can be utilized for a real electron beam energy spread cooling in X-ray FEL linear accelerators [17].

In this paper, systematical studies for the PEHG have been presented. The principle of the PEHG is introduced in Sec. II. Analytical estimates and 1D simulation results are given in Sec. III to present the physical mechanism of the phase-merging effect and the possibility of imprinting ultra-high harmonic microbunching into the electron beam with a relatively small energy spread using this technique. Sec. IV gives an optimized design for a soft x-ray FEL with realistic parameters based on the PEHG. Some practical constraints that may deteriorate the performance of PEHG are studied in Sec. V. Finally, we conclude in Sec. VI.

## 2. Principles of PEHG

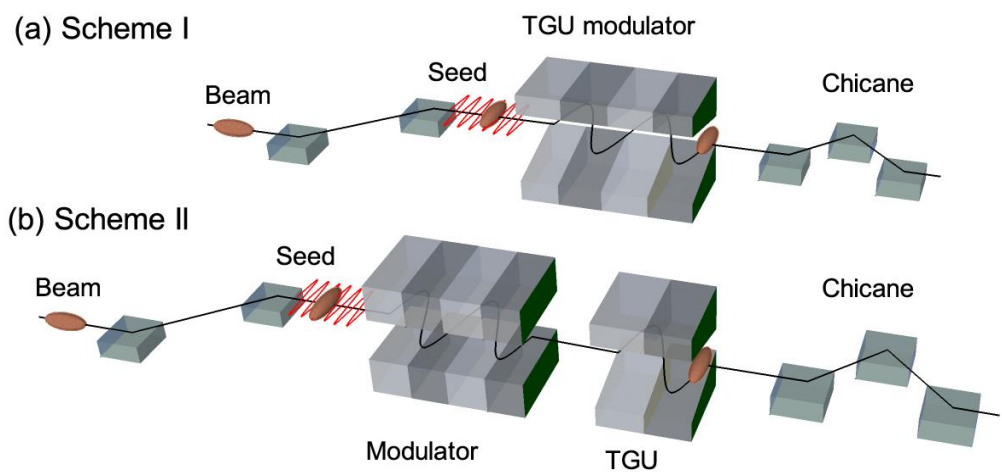

**Figure 1.** (a) Original PEHG scheme with a TGU modulator for energy modulation and phase-merging simultaneously; (b) An PEHG variant with a normal modulator for energy modulation and a TGU for introducing the phase-merging effect.

The initial proposed PEHG consists of a dogleg followed by a HGHG configuration with a TGU modulator, as shown in Fig. 1 (a). The dogleg with dispersion $\eta$ is used to transversely disperse the electron beam, while the TGU modulator is used for the energy modulation and precisely manipulating the electrons in the horizontal dimensional. It is found that these two functions of TGU modulator can be separately performed by employing the scheme II as shown in Fig. 1 (b). The normal modulator is used for the energy modulation and the TGU is only used for transverse manipulation of electrons, which will be much more flexible for practical operation. The TGU in scheme II also can be replaced by other kinds of devices with transverse gradient magnet field, e.g. particularly designed wigglers or small chicanes. For the convenience of theoretical analysis, we consider the scheme II first, and then promote the conclusions to scheme I.

Following the notation of ref. [15], we also assume an initial Gaussian beam energy distribution with an average energy $\gamma_0 mc^2$ and use the variable $p = (\gamma - \gamma_0)/\sigma_\gamma$ for the dimensionless energy deviation of a particle, where $\sigma_\gamma$ is the rms energy spread. So the initial longitudinal phase space distribution should be $f_0(p) = N_0 \exp(-p^2/2)/\sqrt{2\pi}$. Assuming the initial horizontal rms beam size is $\sigma_x$ and use $\chi = (x - x_0)/\sigma_x$ for the dimensionless horizontal position of a particle, then the horizontal electron beam distribution can be written as $g_0(\chi) = N_0 \exp(-\chi^2/2)/\sqrt{2\pi}$. After the dogleg, $\chi$ is changed to

$$\chi' = \chi + Dp, \qquad (1)$$

where $D = \eta \sigma_\gamma / \sigma_x \gamma$ is the dimensionless strength of the dogleg, and the horizontal beam distribution becomes

$$g_1(p, \chi) = \frac{N_0}{\sqrt{2\pi}} \exp\left[-\frac{1}{2}(\chi - Dp)^2\right], \qquad (2)$$

After passage through the modulator, the electron beam is modulated with the amplitude $A = \Delta\gamma/\sigma_\gamma$, where $\Delta\gamma$ is the energy modulation depth induced by the seed laser, and the dimensionless energy deviation of the electron beam becomes $p' = p + A\sin(k_s z)$, where $k_s$ is the wave number of the seed laser. The two-dimensional distribution function after the interaction with the seed laser can be written as

$$h_1(\zeta, p, \chi) = \frac{N_0}{2\pi} \exp\left[-\frac{1}{2}(p - A\sin\zeta)^2\right] \exp\left\{-\frac{1}{2}\left[\chi - D(p - A\sin\zeta)\right]^2\right\}, \qquad (3)$$

where $\zeta = k_s z$ is the phase of the electron beam. Then sending the electron beam through a TGU with transverse gradient $\alpha$ and central dimensionless parameter of $K_0$ converts the longitudinal coordinate z of electrons with different horizontal position into

$$z' = z + \frac{L_m K_0^2}{2\gamma^2}\left[\frac{\alpha\chi'}{\sigma_x} + \frac{1}{2}\left(\frac{\alpha\chi'}{\sigma_x}\right)^2\right], \qquad (4)$$

where $L_m$ is the length of TGU. Considering that the transverse electron beam size is usually quite small for FEL, Eq. (4) can be re-written as

$$z' = z + \frac{L_m K_0^2 \alpha \sigma_x}{2\gamma^2} \chi', \qquad (5)$$

and this makes the electron beam distribution after TGU become:

$$h_2(\zeta, p, \chi) = \frac{N_0}{2\pi} \exp\left\{-\frac{1}{2}\left[p - A\sin(\zeta - T\chi)\right]^2\right\} \exp\left[-\frac{1}{2}\left\{\chi - D\left[p - A\sin(\zeta - T\chi)\right]\right\}^2\right], \qquad (6)$$

where

$$T = \frac{k_s L_m K_0^2 \alpha \sigma_x}{2\gamma^2} \qquad (7)$$

is the dimensionless gradient parameter of the TGU. After passing through the DS with the dispersive strength of $R_{56}$, the longitudinal beam distribution evolves to

$$h_{PEHG}(\zeta, p, \chi) = \frac{N_0}{2\pi} \exp\left\{-\frac{1}{2}[p - A\sin(\zeta - T\chi - Bp)]^2\right\} \exp\left[-\frac{1}{2}\{\chi - D[p - A\sin(\zeta - T\chi - Bp)]\}^2\right], \quad (8)$$

where $B = R_{56} k_s \sigma_\gamma / \gamma$ is the dimensionless strength of the DS. Integration of Eq (8) over $p$ and $x$ gives the beam density $N$ as a function of $\zeta$, $N(\zeta) = \int_{-\infty}^{\infty} dx \int_{-\infty}^{\infty} dp\, h_{PEHG}(\zeta, p, x)$. And the bunching factor at $n$th harmonic can be written as

$$b_n = \frac{1}{N_0} \int_{-\infty}^{\infty} dp\, e^{-inp(TD+B) - inT\chi} f_0(p) g_0(\chi) \langle e^{-in(\zeta + AB\sin\zeta)} \rangle = J_n[nAB] e^{-(1/2)[n(TD+B)]^2} e^{-(1/2)(nT)^2}, \quad (9)$$

For the case without TGU, i.e. $T = 0$, Eq. (9) reduces to the well-known formula for the bunching factor in a standard-HGHG FEL.

For the harmonic number $n > 4$, the maximal value of the Bessel function in Eq. (9) is about $0.67/n^{1/3}$ and is achieved when its argument is equal to $n + 0.81 n^{1/3}$. For a given value of energy modulation amplitude $A$, the optimized strength of the DS should be

$$B = (n + 0.81 n^{1/3})/nA. \quad (10)$$

The maximal value of Eq. (9) will be achieved when $TD = -B$, which gives the optimized relation of $\alpha$ and $\eta$:

$$\alpha\eta = -\frac{2\gamma(n + 0.81 n^{1/3})}{nAk_s L_m K_0^2 \sigma_\gamma} \quad (11)$$

Noticing that the third term in the right hand of Eq. (9) can be quite close to one when adopting a large $A$ and $\eta$ or a small horizontal beam size $\sigma_x$, so the maximal value of the nth harmonic bunching factor for PEHG will approach

$$b_n \approx 0.67/n^{1/3}, \quad (12)$$

which is much larger than that of a standard-HGHG.

For the scheme I as shown in Fig. 1(a), the energy modulation process and the phase-merging process are accomplished simultaneously when the electron beam passes through the TGU modulator. The electron relative phase advance caused by the gradient of the TGU is the same for scheme I and scheme II. However, a factor of 1/2 should be introduced in the right hand of Eq. (7), because the energy modulation approximately increase linearly with the modulator period number $N_m$ thus the phase advance obtained by integration over the modulator length contributes a factor of 1/2. So we get the optimized dimensionless gradient parameter of TGU should become $TD = -2B$, and the relation between $\alpha$ and $\eta$ for scheme I becomes:

$$\alpha\eta = -\frac{4\gamma(n + 0.81 n^{1/3})}{nAk_s L_m K_0^2 \sigma_\gamma} \quad (13)$$

It is precisely consistent with the earlier results given in ref. [16].

## 3. Physical mechanism of PEHG

The physical mechanism behind the PEHG is the transverse-longitudinal phase space coupling. The evolution of beam longitudinal phase space is illustrated in Fig. 2 for the scheme II in Fig. 1. For simplicity, here we assume the horizontal beam size $\sigma_x = 0$ and only show the phase space within one seed wavelength region. The energy modulation amplitude is chosen to be $A = 3$ here, and the optimized condition for the 50th harmonic bunching is $B = -TD \approx 0.35$ according to Eq. (10).

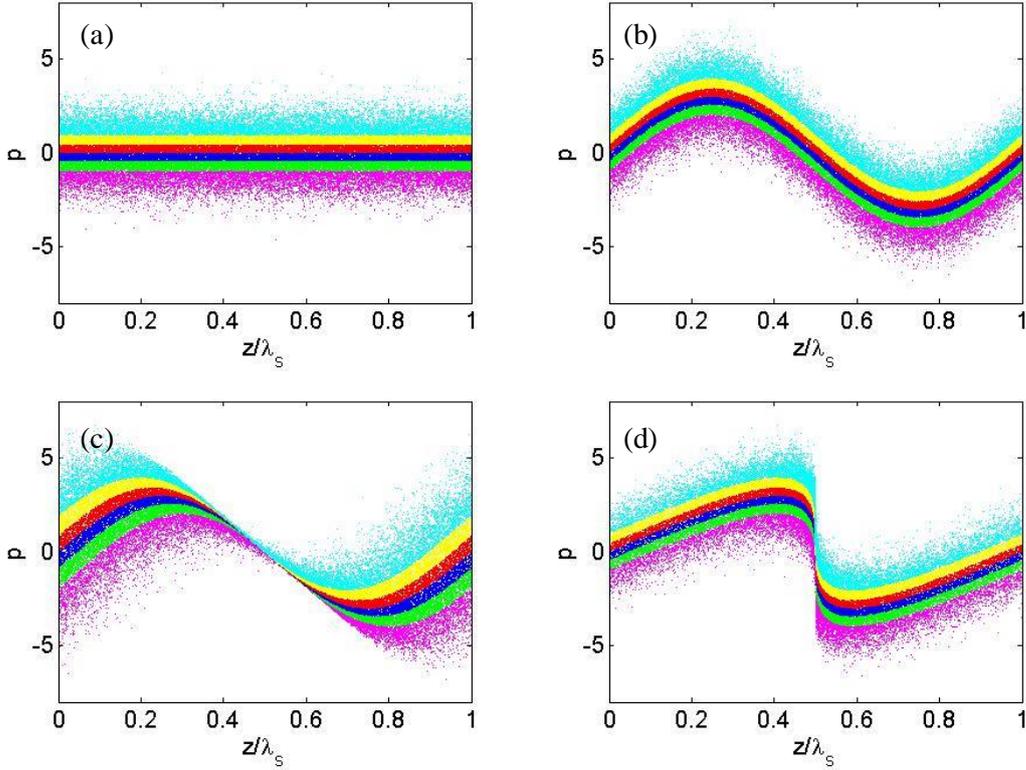

**Figure 2.** Longitudinal phase space evolution in scheme II: (a) initial phase space after passing through the dogleg; (b) phase space at the exit of the conventional modulator; (c) phase space at the exit of the TGU; (d) phase space at the exit of the DS.

The initial longitudinal phase space after passage through the dogleg is shown in Fig. 2 (a), where different colors represent for different regions of beam energy and so also represent for different horizontal positions of the electrons with respect to the reference electrons with central beam energy. After interaction with the seed laser in the conventional modulator, the longitudinal phase space of the beam evolves to that shown in Fig. 2(b). The strong optical field induces a rapid coherent growth of the electron beam energy spread. When the beam travels through the TGU, electrons with different colors (different transverse position) will meet different undulator $K$ values, thus result in the different travel path lengths in TGU. By properly choosing the gradient of TGU according to Eq. (11), the phase space will evaluate to that in Fig. 2(c). The electron energy is unchanged during this process. However, the electrons with the same energy will merge into a same longitudinal phase around the zero-crossing of the seed laser due to the relative phase shift of the electrons in TGU. This phenomenon is what we called the "phase-merging effect". After passage through TGU, electrons enter the dispersion section where the beam phase space is rotated and the bunching at the desired harmonic is optimized, as shown in Fig. 2(d). One can find that most of the electrons are compressed into a small region around the zero-phase, which indicates that the density modulation has been significantly enhanced for high harmonics.

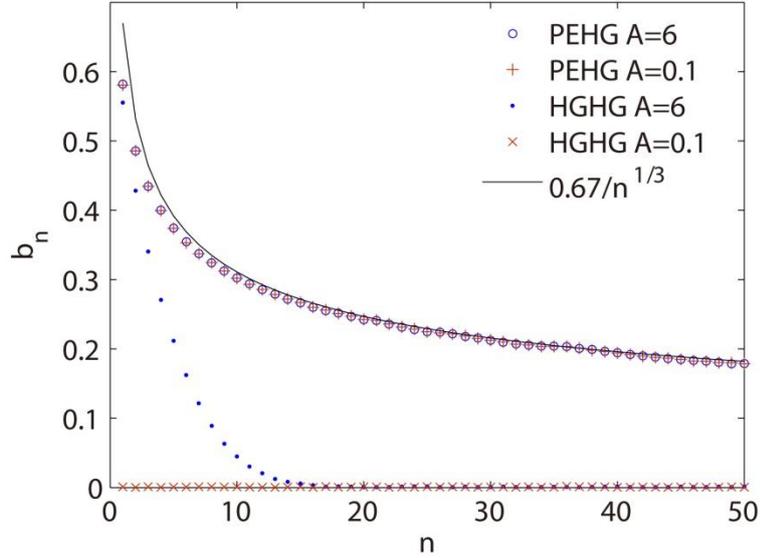

**Figure 3.** Comparison of the bunching factor of PEHG and standard HGHG with different energy modulation amplitudes. The black line is the theoretical prediction of the maximal bunching factor.

It can be deduced form Eq. (9) that the maximal bunching factor of PEHG is mainly determined by the Bessel function term and has little dependence on the absolute value of $A$ when $\sigma_x$ is small or $\eta$ is quite large. Fig. 3 shows the simulation results of the maximal bunching factor distributions of PEHG for different energy modulation amplitudes under the condition of $\sigma_x = 0$. For comparison purpose, the bunching factor distributions for the optimized standard HGHG with the same energy modulation amplitudes are also shown. One can clearly see that the bunching factor exponentially decreases as the harmonic number increases for standard HGHG. However, for PEHG, the bunching factor decreases as $n^{-1/3}$ and the maximal value fit quite well with the theoretical prediction curve for $n > 4$.

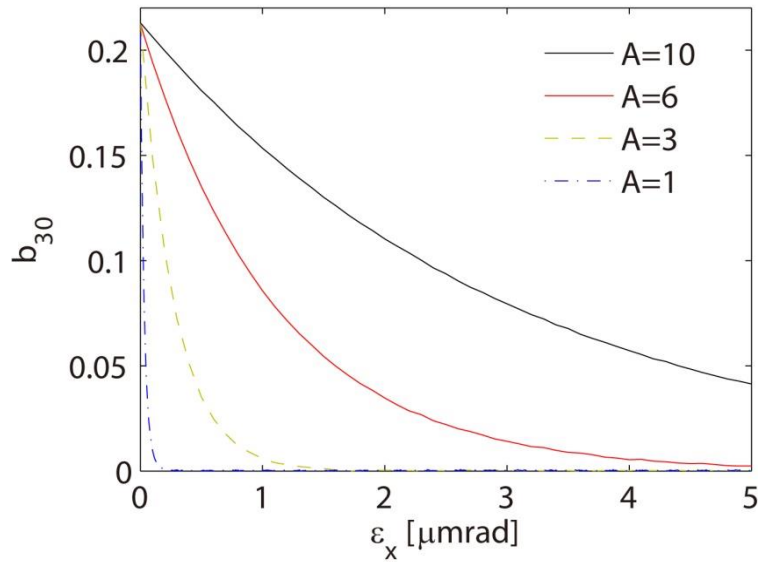

**Figure 4.** The 30th harmonic bunching factor of PEHG as a function of the horizontal emittance for different energy modulation amplitudes.

For a realistic electron beam, the intrinsic horizontal beam size $\sigma_x$ cannot be neglected. It will induce an effective energy spread into the electron beam because of the transverse field gradient of TGU. The effective energy spread can be written as [18]

$$\sigma_{eff} = \sigma_x / \eta. \quad (13)$$

Using the optimized condition of PEHG: $T\eta\sigma_\gamma / \sigma_x\gamma = -B$, plug Eq. (13) into Eq. (9), and we arrive

$$b_n = J_n[nAB]e^{-(1/2)(nk_sR_{56}\sigma_{eff})^2}. \quad (14)$$

One may found that the bunching factor formula of Eq. (14) is reduced to the bunching factor form of a standard HGHG. The only difference is that the initial beam energy spread has been replaced by $\sigma_{eff}$. Here we define an energy spread compression factor $C = \eta\sigma_\gamma / \gamma\sigma_x$, which can be used to measure the phase-merging effect. The intrinsic beam size is determined by the normalized horizontal emittance $\varepsilon_x$ and the beta function $\beta$. For a relatively short modulator of length $L_m$, it is reasonable to take $\beta \approx L_m / 2$, and hence $\sigma_x = \sqrt{\varepsilon_x L_m / 2\gamma}$. By using the realistic parameters of Shanghai Soft X-ray FEL (SXFEL) project [19], Fig. 4 shows the 30[th] harmonic bunching factor as a function of the initial horizontal emittance. The beam energy is 840MeV with energy spread of about 100 keV, the dispersion of the dogleg is $\eta = 1m$, the gradient of the TGU is $\alpha = 20\ m^{-1}$, and the average beta function in the short modulator is $\beta = 0.5m$. The energy modulation amplitude has been changed from 1 to 10. One can found from Fig. 4 that the bunching factor decreases quickly as the horizontal emittance increases when $A$ is smaller than 3. However, the bunching factor is still acceptable for $\varepsilon_x = 1\mu mrad$ when $A$ is larger than 6. For the case of $\varepsilon_x = 1\mu mrad$ and $A = 6$, the comparison of the bunching factor of PEHG and HGHG is shown in Fig. 5. The energy spread compression factor is calculated to be $C \approx 5.74$ or this case, which approximately makes the harmonic number increase 6 times with the same bunching factor for high harmonics.

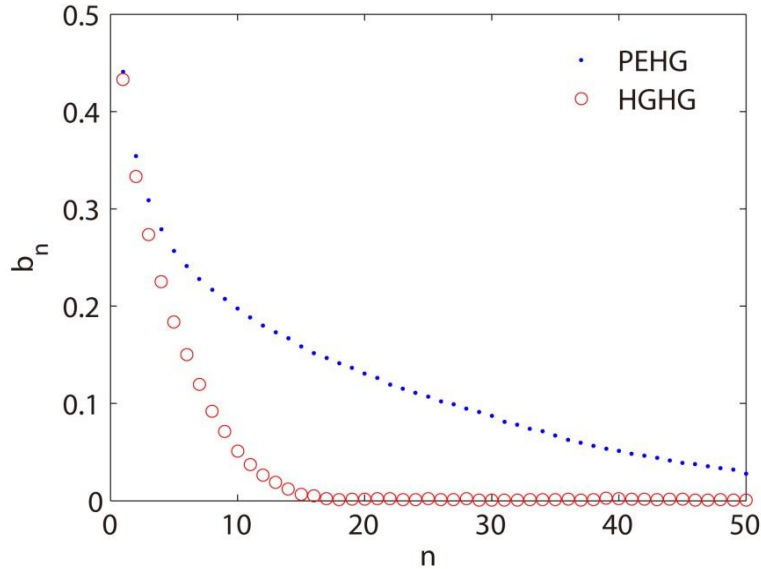

**Figure 5.** Comparison of the bunching factor of PEHG and standard HGHG with realistic parameters.

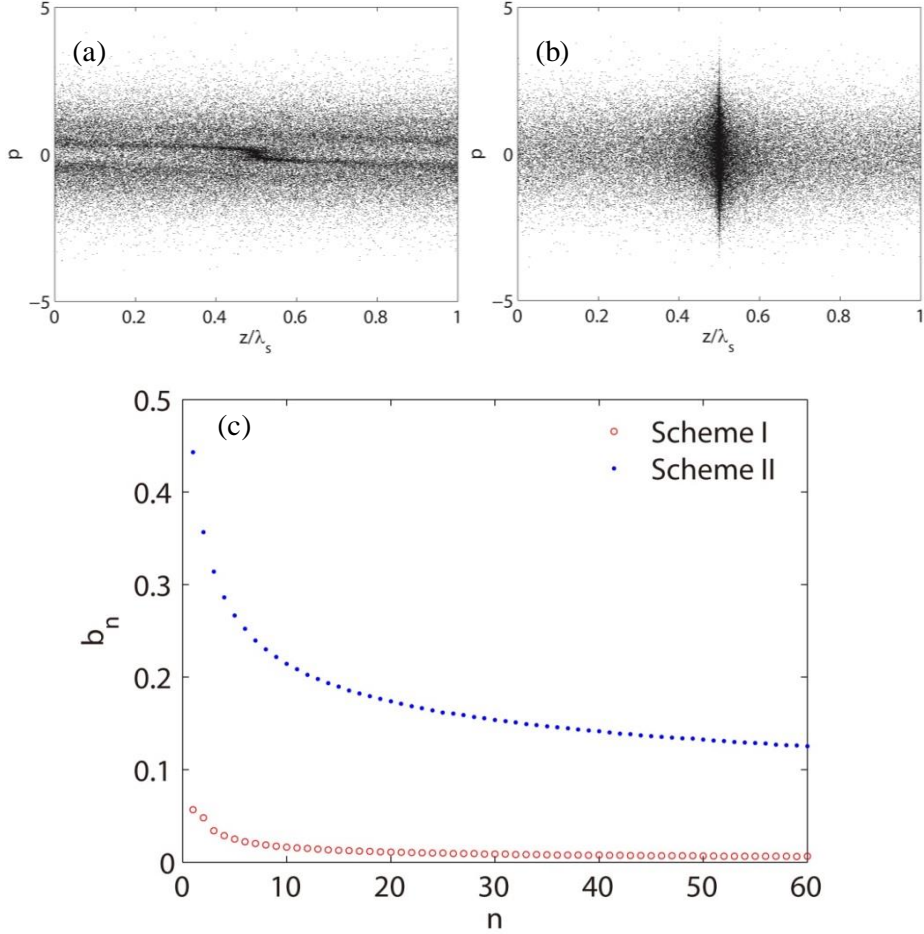

**Figure 6.** The longitudinal phase spaces of the electron beams for scheme I (a) and scheme II (b) in Fig. 1 at the entrance to the radiators and corresponding bunching factor distributions (c).

It should be pointing out here that the bunching factor are nearly the same for the two schemes shown in Fig. 1 when $A$ is much larger than 1. However, for a relatively small $A$, the final longitudinal phase space distribution and the bunching factor will be quite different for these two schemes. Fig. 6 shows the simulation results for the two schemes when $A = 0.1$ and $\varepsilon_x = 0\,\mu mrad$. According to the non-linear effect during the modulation process for scheme I, the bunching factor decreases for all harmonics. However, the bunching factor is nearly the same to the $A = 3$ case as shown in Fig. 3 for scheme II, which demonstrates the theoretical prediction.

## 4. Generation of soft x-ray radiation

To illustrate a possible application with realistic parameters and show the parameter optimization method of PEHG, we take the nominal parameters of the SXFEL. The SXFEL test facility aims at generating 8.8 nm FEL from a 264 nm conventional seed laser through a two-stage cascaded HGHG. The electron beam energy is 840 MeV with slice energy spread of about 100 keV. The beam peak current is over 600 A. As mentioned above, the bunching factor of PEHG is quite sensitive to the beam emittance. The optimized 30$^{th}$ harmonic bunching factor and 3D gain length of the 8.8 nm radiation as a function of the initial horizontal emittance are shown in Fig. 7. From Fig. 7 (a), one can find that the bunching factor decrease quickly as the emittance increase when $\eta$ is smaller than 0.5 m, and the bunching factor can be well maintained for $\eta>1$ m. However, when the dispersion is too large, it will contribute to a FEL gain reduction due to the increased beam size. In order to ensure adequate gain in the radiator, the dispersion induced beam size is required to be not larger than the intrinsic horizontal beam size contributed by the radiator beta function. For SXFEL, the beam size in the radiator is about 100 μm level, considering the beam energy spread of 100 keV, the maximum dispersion permitted is about 1 m. The 3D FEL gain length as a function of horizontal emittance with different dispersion are calculated and shown in Fig. 7 (b). The gain length can be well controlled in 2 m for 1 μmrad emittance and 1 m dispersion case, which is reasonable for a seeded soft x-ray FEL. With the above parameters, the horizontal beam size will be increased from 100 μm to about 220 μm when η is turned from 0 to 2 m. As the transverse beam size should be calculated by $\sigma = \sqrt{\sigma_x^2 + \sigma_y^2}$, where the $\sigma_y^2$ is unchanged for different η, the 3D gain length will change little (form 1.5 m to 1.8 m) when η is smaller than 1 m.

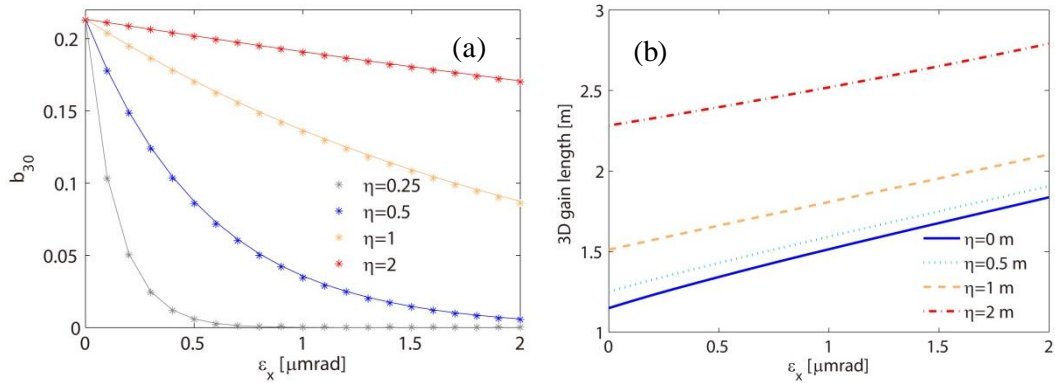

**Figure 7.** (a) The 30$^{th}$ harmonic bunching factor and (b) three-dimensional gain length of PEHG as a function of the horizontal emittance for different dispersion strength.

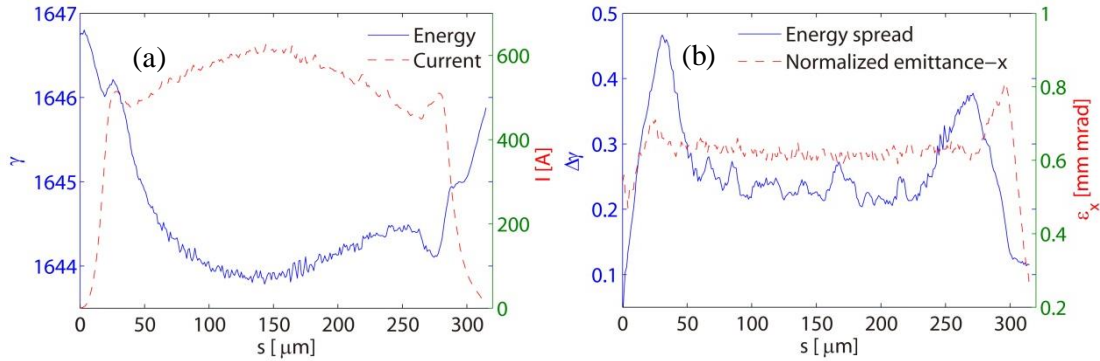

**Figure 8.** Simulated parameters at the exit of the linac. (a) Beam energy and current distribution along the electron beam. (b) Slice energy spread and normalized emittance distribution along the electron beam.

With the above parameters, start-to-end tracking of the electron beam, including all components of SXFEL, has been carried out. The electron beam dynamics in photo-injector was simulated with ASTRA [20] to take in to account space-charge effects. ELEGENT [21] was then used for the simulation in the remainder of the linac. The slice parameters at the exit of the linac are summarized in Fig. 8. The beam energy in the central part of the electron beam is around 840 MeV and the peak current is about 600 A. A constant profile is maintained in the approximately 600 fs wide and over 500 A region. A normalized emittance of approximately 0.65 μmrad and slice energy spread of about 100 keV are observed in Fig. 8 (b). Fig. 9 shows the transverse beam central and beam size changes after passage through the dogleg. The average value of the horizontal beam size $\sigma_x$ is increased from about 60 μm to about 70 μm, which will not significantly affect the FEL performance. However, the horizontal beam position is changed a lot due to the large energy chirp in the electron beam as shown in Fig. 9 (b).

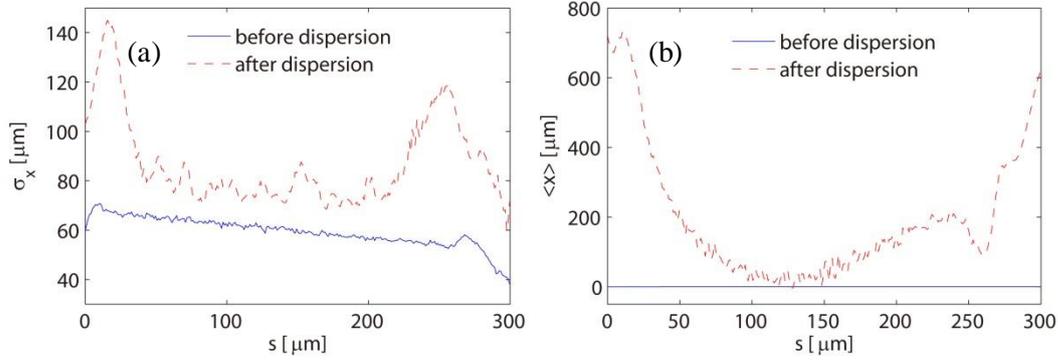

**Figure 9.** Comparisons of the horizontal beam size (a) and beam central position change before and after the dogleg in the simulation.

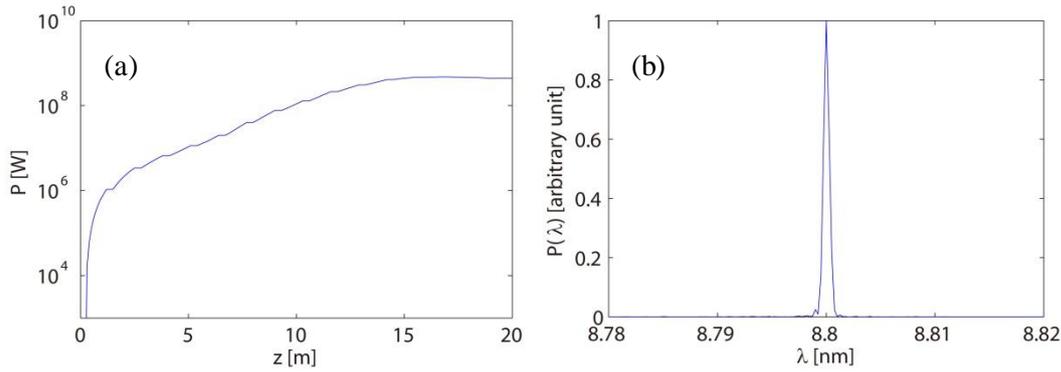

**Figure 10.** FEL performance of PEHG at 8.8 nm. (a) Radiation peak power as a function of undulator distance. (b) Radiation spectrum at saturation.

The FEL performance of PEHG was simulated by the upgraded three-dimensional FEL code GENESIS [22] based on the output of ELEGENT. A 265 nm seed pulse with longitudinal pulse length much longer than the bunch length is adopted in the simulation. The length of TGU modulator is about 1 m with period length of 80 mm and K value of around 5.8. To maximize the bunching factor at 30$^{th}$ harmonic of the seed laser, the optimized parameters are set to be $A = 5, B = 0.2, \alpha = 50\ m, \eta = 0.5\ m$. The bunching factor at the entrance of the radiator is about 5.6%, which fit quite well with the theoretical prediction of Fig. 7 (a). The period length of the radiator is 25 mm with K value of about 1.3. The evolution of the radiation peak power is shown in Fig. 10. The large bunching factor at the entrance to the radiator offered by the PEHG scheme is responsible for the initial steep quadratic growth of the power. The significant enhancement of the performance using the PEHG is clearly seen in Fig. 10 (a) where the peak power of the 30$^{th}$ harmonic radiation exceeds 400MW, which is quite close to the output peak power of the original design of SXFEL with two-stage HGHG. Moreover, the 8.8nm radiation saturates within 15 m long undulator, which is in the range of original design of SXFEL. The single-shot radiation spectrum at saturation is shown in Fig. 10 (b), from which one can find that the bandwidth of the radiation at saturation is quite close to transform-limited. Further studies show that the output spectrum of the PEHG is immune to the residual beam energy chirp due to the transverse gradient.

### 5. Some practical issues

The unique feature of PEHG is utilization of a TGU device with transverse gradient of $\alpha$. We will discuss in this section some practical issues that may affect the performance of PEHG.

Unlike a conventional planar modulator undulator in the standard-HGHG, the TGU in PEHG will introduce an external focusing in the transverse dimension due to the gradient field, which will results in the deviation of the electron trajectory in horizontal. For a TGU, the magnetic field distribution can be written as

$$B_y(x,z) = B_0(1+\alpha x)\sin k_u z \quad (15)$$

where $B_0$ is the undulator peak magnetic field. The electric field of the seed laser can be simply represented as

$$E_x(z) = E_0 \sin(k_s z + \varphi_0) \quad (16)$$

where $E_0$ is the peak electric field and $\varphi_0$ is the carrier envelop phase of the seed laser. Then the trajectory equations of electron with initial horizontal position $x_0$ can be written as

$$\frac{dx}{dz} = \frac{dx}{cdt} \quad (17)$$

$$\frac{d^2 x}{dz^2} = -\frac{e}{\gamma mc^2}\cdot\frac{dx}{dt}\cdot B_0(1+\alpha x)\sin k_u z = -\frac{e}{\gamma mc}\cdot\frac{dx}{dz}\cdot B_0\left[1+\alpha\left(x_0+\frac{K_0}{k_u\gamma}\sin k_u z\right)\right]\sin k_u z$$

$$= -\frac{e}{\gamma mc}\cdot\frac{dx}{dz}\cdot B_0\left[(1+\alpha x_0)\sin k_u z + \frac{\alpha K_0}{k_u\gamma}\sin^2 k_u z\right] \quad (18)$$

The x-deviation after passage through the modulator is

$$\Delta x = \frac{\alpha}{4\gamma^2}K_0^2 N_m^2 \lambda_u^2 \quad (19)$$

According to the parameters used in Sec. V, $\Delta x$ is calculated to be about 96 μm, which will not significantly affect the performance of PEHG. It can be found from Eq. (17) that the x-deviation can be compensated by introducing an external magnetic field:

$$B_{external} = \frac{\alpha K_0}{2k_u\gamma}B_0 \quad (20)$$

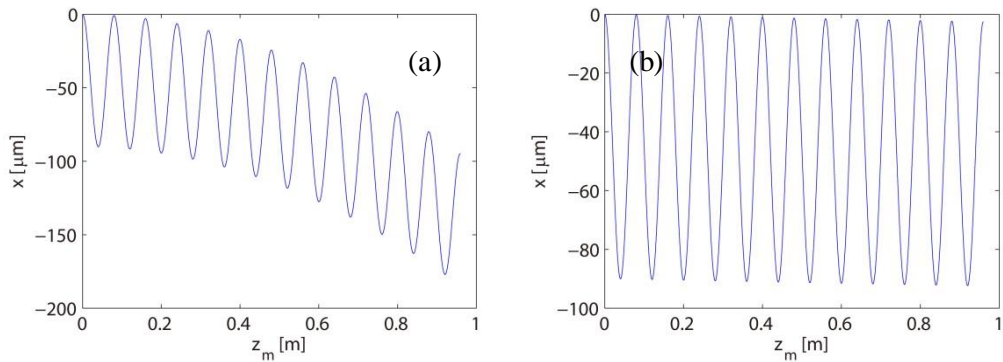

**Figure 11.** Electron trajectory in horizontal dimension: (a) TGU modulator; (b) TGU with correction magnetic field.

To illustrate the particle trajectory in the TGU and check the simulation results of GENESIS, we develop a three-dimensional algorithm based on the fundamentals of electrodynamics when considering the appearance of gradient undulator magnetic field and laser electric field in the time domain [23]. The simulation results are shown in Fig. 11. It can be found from Fig. 11 (a) that the deviation of the electron trajectory in horizontal at exit of the modulator is about 95 μm, which fit quite well with the theoretical calculation. As the transverse laser size is much larger than the beam size, i.e. about 1000 μm (rms) in this simulation, the horizontal deviation will not significantly affect the modulation process. For the case that shown in Fig. 11 (a), the bunching factor decreases from about 5.6% to about 5.4%. And this deviation can be compensated by introducing an external magnetic field as shown in Fig. 11 (b), the magnetic field is about 5.6Gs.

The sensitivity of the bunching factor to the shot-to-shot fluctuations of the laser power has also been studied by introducing random fluctuations of the laser power in the modulator within ±5%. The resulting 1000 shots of fluctuations of the 30th harmonic bunching factor are shown in Fig. 12. One can find that, with 5% tolerance on the seed laser peak power, the bunching factor of PEHG can be well maintained over 5%.

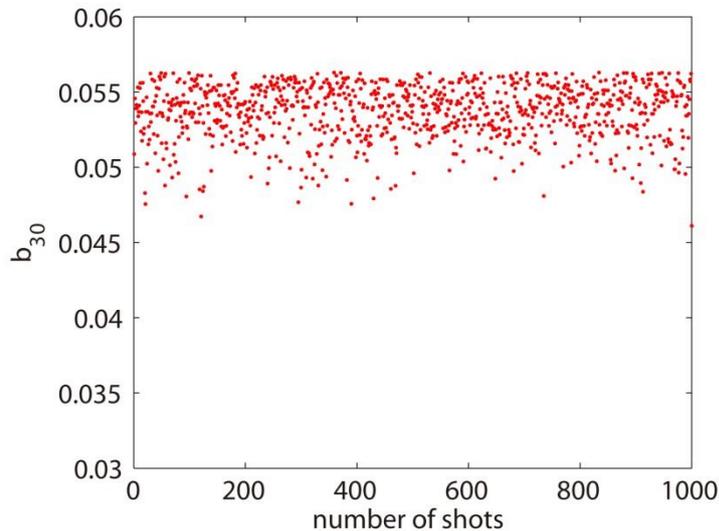

**Figure 12.** The 30$^{th}$ harmonic bunching factors at various shots for a fluctuating amplitude of the seed laser power

## 6. Conclusion

In summary, intense analytical and numerical investigations of the PEHG schemes have been accomplished. The results demonstrate the potential of generating ultra-high harmonic radiation with a relatively small energy modulation by a single stage PEHG. It is found that, the optimized $n^{th}$ harmonic bunching factor of PEHG is nearly only determined by the maximal value of the $n^{th}$ order Bessel function, which decreases as $n^{-1/3}$. The transverse dispersion induced beam size increase will not degrade the FEL performance when the system parameters are properly set. For PEHG FEL operated at 8.8nm directly from 264nm, the numerical example demonstrates a peak power exceeding 400MW, which is comparable with that of the original two-stage HGHG design. Considering that the ability of exploiting the full electron bunch in the PEHG, the output bandwidth and the pulse energy will be significantly improved, and thus leads a FEL average brightness 2 orders of magnitude higher than the two-stage HGHG baseline.

In addition to generation of fully coherent radiation at soft x-ray, the concept of the phase-merging effect also offers a novel method for flexibility beam energy spread control, which may be useful for cooling electron beam energy spread or ultra-intense and ultra-short FEL pulses generation.

**Acknowledgments**
The authors would like to thank B. Liu, T. Zhang, G. Stupakov, Y. Ding, D. Xiang and Z. Huang for helpful discussions and useful comments. This work is supported by the Major State Basic Research Development Program of China (2011CB808300) and the National Natural Science Foundation of China (11175240, 11205234 and 11322550).